# Entity Based Peer to Peer in a Data Grid Environment

B. Hudzia, L. McDermott, T. N. Ellahi, T. Kechadi
Parallel Computational Research Group, Department of Computer Science
University College Dublin, Belfield,
Dublin 4, Ireland
E-mail :{ benoit.hudzia, liam.mcdermott, tariq.ellahi, tahar.kechadi} @ucd.ie

**Abstract**

During the last decade there has been a huge interest in Grid technologies, and numerous Grid projects have been initiated with various visions of the Grid. While all these visions have the same goal of resource sharing, they differ in the functionality that a Grid supports, the grid characterisation, programming environments, etc. In this paper we present a new Grid system dedicated to deal with data issues, called DGET (Data Grid Environment and Tools). DGET is characterized by its peer-to-peer communication system and entity-based architecture, therefore, taking advantage of the main functionality of both systems; P2P and Grid. DGET is currently under development and a prototype implementing the main components is in its first phase of testing. In this paper we limit our description to the system architectural features and to the main differences with other systems.

## I. INTRODUCTION

In recent years, Internet-scale systems have been developed and deployed to share resources at a very large scale across the traditional organizational boundaries. The need for constructing such systems was motivated by the increasingly complex requirements of the modern applications from diverse disciplines. Such global scale systems provide opportunities to harness idle resources which are distributed and heterogeneous. Another benefit offered by such systems is that they allow coordinated use of resources from multiple organisations. Thus, these wide-area systems may span multiple organisations and form virtual organisations on top of the existing organizational hierarchies.

Two such systems exploiting these views include Grid systems and Peer-to-Peer (P2P) systems. Grid and P2P have seen a rapid evolution and widespread deployment. The two technologies appear to have the same final objective, pooling and coordinating large sets of distributed resources [FOS 03]. During the last few years various projects have been undertaken to try to merge the two complementary approaches of these technologies, such as NaradaBrokering [PAL 03]. Also various modifications to the Globus toolkit [FOS 97] have been proposed to include P2P technology and thus improving the discovery system [Talia 03].

Usually Grid systems were designed to run applications with intensive computing and storage needs across the traditional organisational boundaries [FOS 01, FOS02, CHE 99]. Grid systems are characterised by sophisticated resource management and data transfer components. P2P systems on the other hand were mainly designed for resource sharing, mostly files. Therefore, the focus of P2P systems is on providing sophisticated resource discovery capabilities. Both approaches have their own advantages and disadvantages.

In this paper we present the DGET (Data Grid Environment and Tools) project undertaken by the Parallel Computational Research Group at UCD. DGET is a grid computing system. The two main distinguishing features of DGET are its peer-to-peer communication system and its entity based nature. The entity concept is very much like an agent in the sense of its autonomy and dynamicity, but it has features which make it different from agents. The idea of using such a concept is not new. There has already been much talk of integrating agent based systems and grid systems [FOS 04], allowing each system to make use of the features provided by the other. The dynamic nature of agents allows them to be deployed with little administration while grid systems provide a more predictable and stable system. The DGET Entity is based on the same principle.

The rest of the paper is structured as follows. Section 2 gives a detailed introduction of the DGET system. Section 3 explains the entity concept. Section 4 explains the communication subsystem in DGET and Section 5 is related work, where we compare our approach with other systems. Section 5 concludes with the current status of the project and future work.

## II. DGET SYSTEM

There are a few systems that combined the concepts from both Grid and P2P systems. Such hybrid systems are called P2P Grids. DGET adopts the same approach and exploits the advantages of both systems. DGET has the following distinguishing features.

*Transport independent communication:* The first distinguishing feature of DGET is its transport independent communication subsystem. Applications use a transport independent interface exposed by DGET. No changes need to be made when a different transport protocol is used in the underlying system.

*Decentralized Resource discovery:* DGET adopts the decentralized resource discovery approach borrowed from the P2P systems. No centralized servers are maintained to hold information about the resources in the system. This results in improving the scalability of the system. P2P style resource discovery in Grid systems have been investigated in [IAM01, SCH 03, XU 02]. DGET improves the decentralized P2P style resource discovery approach by introducing the notion of customized neighbourhoods [ELL 04]. This customized neighbourhood helps reduce resource discovery time.

*Uniform resource Interfaces:* Resources in DGET systems are represented through a standard and uniform interface. This approach helps in masking the intra-resource heterogeneity. Users don't have to master the entire heterogeneous interface. New resources can be seamlessly added to the system.

*Fine grained security:* DGET provides mechanisms to specify and enforce fine grained authorization policies on the resources.

*Minimum administration overhead:* DGET, contrary to the existing grid systems, doesn't have a great deal of administration overhead. No complex setup is required for the deployment of the DGET system.

*Self-organization:* DGET doesn't have a static topology, unlike current grid systems. DGET adopts a self-organizing topology. Nodes in the system don't have to be statically setup in some topology. The topology of the system evolves as a result of interactions between the nodes of the system. Nodes join and leave the system with minimum overhead incurred by the system.

*Entity based system:* An entity is any physical or logical object which has the capability of performing some functionality. Examples of entities can be a Processor entity that executes applications, a storage entity that provides data storage functionality, a data mining entity providing the data mining capability, etc.

The DGET system is composed of several elements as shown in Figure 1. The entities sit on top of the nucleus; provide high-level functions of the system, and handle user P2P and/or Grid applications. Like in most systems, the DGET nucleus provides low-level systems functions, however it also implements a very advanced communication system.

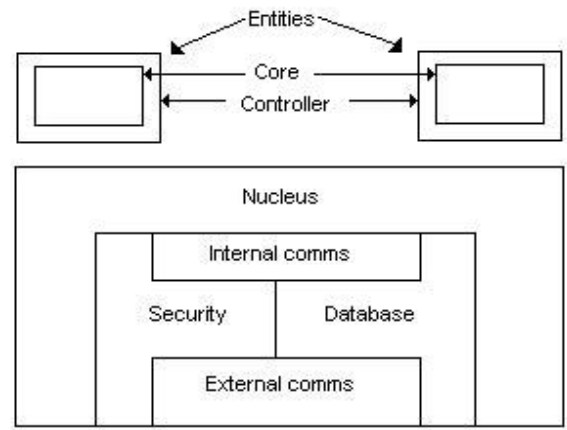

Figure 1: DGET System Architecture

There has been a distinction made between internal and external communication, but both are encapsulated within the nucleus itself. Internal communication is when two entities communicating are on the same nucleus and it boils down to taking a message from one entities message queue and adding it to another entities message queue. External communication is obviously more complicated as it requires communication between different nuclei which are most likely on different physical machines. The communication system will be explained in more detail in Section 4.

The entities are also composed of two elements called the Entity Core and the Entity Controller. The Entity Controller is a runnable object, i.e. a thread, which interacts with the nucleus, passing entities messages and providing any other necessary management functions. The Entity Core provides the functionality of the entity. This is in the form of another object, but unlike the Entity Controller the Entity Core object is not runnable, it runs within the Entity Controller thread. This system allows functionality to be added to the system in the form of different Entity Cores while the same Entity Controller type is used. The entity system will be explained in more detail in the Section 3.

*A. System entities*

System entities represent the core of the system. They implement the various protocols required to provide the different functionality needed for the p2p/grid system. These entities rely heavily on the uniform semantics we define for DGET system. The main entities are:
*Security entity*, applying local policies to a group of entities belonging to a nucleus or a group of nuclei.
*Nucleus entity manager*, this entity represents the interface between the nucleus and the physical system supporting the nucleus. This entity will ensure that the various entities run smoothly with each other on a localized scale.

**Resource discovery entity** provides p2p like protocols for searching, discovering resources and routing between various entities.

*B. The Nucleus*

The nucleus represents the kernel of the DGET system. It provides basic security and a complete communication system for all local and remote communicating entities as well as providing various other critical functions. The nucleus is composed of four main components: internal transport layer, external transport layer, basic security, and database (see Figure 1).

### III. DGET ENTITY MODEL

A DGET Entity is the conceptual abstraction of all the objects existing in the system. All the objects whether physical, logical or a combination of both are abstracted and represented as entities.

The DGET entity model is not limited to only the entity structure but also to different categories of entities which can co-exist in the system. Since all these categories have specific and different characteristics we need to come up with a unified interface, which constitutes the entity model. The entity users (which can be other entities) will use the interface to access and manage the entities. The advantage of the entity model is the decoupling of the implementation from the representation. Thus, various implementations can exist in the same DGET system while the same virtual interface can be used to access them in a unified manner. Another advantage of this entity model is the management of entity upgrades. The entity description and representation can change in the entity model but the entity clients need not be upgraded to accommodate this change. Entity clients can still keep using the same virtual interface to access other entities. The main advantage of a virtual entity interface is the uniform access primitives which provides for a diverse range of entity implementations in the system. Thus the heterogeneous nature of resources in the system can be masked making it easier to incorporate and define new types of resources. The clients will only have to learn the virtual interface which can be used to uniformly access every kind of entity. Before going into the details of the virtual interface, a precise definition of an entity is required.

An entity has several features:
- Self describing
- Self contained
- Provides minimal and common interface for all entities
- Has its own agenda
- Can represent a real counterpart, i.e. physical, logical, or both
- Single or composite element
- Can be tightly or loosely coupled with other entities

*A. Entity Components*

In this section we discuss the structural components of any entity in the system. An entity in DGET is composed of two parts, an Entity Core and an Entity Controller. Details of each are given below:

**Entity Core:** This is the part which provides the functional capability of the entity. Typically this part of the entity will be written by an entity developer. The Entity Core will be deployed along with the Entity Controller.

**Entity Controller:** It provides miscellaneous functions to Entity Core including: life cycle management, security, migration, etc. The Entity Controller will be provided by the DGET runtime while deploying the entity. It is at this level that the entity will be abstracted and a uniform access interface will be presented to the entity clients. Thus the heterogeneous implementation details of entities will be masked behind this part of the entity.

*B. Entity Types*

Following on from the brief overview of the entity and its constituent parts, we describe the different types of entities which can be part of the system. Entities can be categorized into two closely related types. The basis of the categorization is the two sets of management operations which can be invoked on entities. These two types are Capability Entities and Activity Entities. Details of these types are given below:

*1) Capability Entity:*

These are the entities which possess the capability to provide some functionality. These are the long-lived entities. Capability entities can be further divided into three types. This division is only taken into account during the creation of new entities. Capability Entity types are explained below:

**Atomic Capability:** Any capability entity which provides a single functionality can be called an Atomic Capability Entity.

**Composite Capability:** A Composite Capability Entity is a higher level entity which may be composed of a number of lower-level atomic or composite entities working together to provide a new capability.
***Aggregate Capability:*** This type of entity is also a multi-capability entity but such an entity is capable of providing all the functionalities individually. While creating such entities the client must specify which capability is needed.

*2) Activity Entity:*

The second type of entity is the Activity Entity. The Activity Entity is the one single instance of the capability entity which will be created to provide functionality to a single user under certain security conditions. In other words, an Activity Entity is a sandboxed version of the Capability Entity which is allowed to use a subset of the Capability Entity's functions. The functions allowed are dependant on the user who created the Activity Entity.

There are a certain set of management operations which can only be invoked on Activity Entities and not on Capability Entities. Such operations include subscription, notification operations, migration operations, etc.

We can give an example to help clarify the differences between the Capability and Activity Entities. Consider a Processor which can be thought of as a Capability Entity. It has the capability to execute application/user jobs. An example of an Activity Entity can be a process launched by a user. The process can consume only a subset of the CPU time depending on the user launching the job. Management operations like, delete/cancel can not be applied to the Processor Entity but they can be invoked on the job, which will result in the deletion of the job from the memory. Another example of the management operation which will make sense only at the activity level is the migrate operation which will migrate the running job to a remote nucleus.

*C. Entity Interfaces*

After explaining the main concepts of the Entity Model we describe the uniform set of management operations which can be applied to all of the entities. Since we have divided entities into two distinct sets based on the management operations that can be invoked on each set, we explain the operations in their respective sets.

*1) Capability Entity Interface:*

***Create:*** This operation can be invoked on any Capability Entity to create an Activity Entity for this Capability. Activity parameters need to be specified to configure the newly created Activity Entity. For aggregate entities offering multiple capabilities, the client also has to specify for what Capability the new Activity should be created. The create operation will return the ID of the newly created Activity Entity. This can then be used to invoke operations on the Activity entity.

***AddCommunityPriviliges:*** This operation will add a new community which can access and use the capability provided by this entity. The parameters to this operation include the community name and set of rights associated with that community.
***UpdateCommunityPriviliges:*** Privileges associated with already added communities can be updated through this operation.
***RemoveCommunityPriviliges:*** Rights for existing communities can be revoked by removing the community from the list of communities allowed to access the capabilities of this entity.

*2) Activity Entity Interface*

***Delete:*** Invoking the delete operation on an Activity Entity terminates the Activity and all the resources allocated to the Activity are reclaimed by the system.

***Subscribe:*** This operation adds the invoking Activity Entity to the list of entities which will be informed when a certain event occurs.
***RemoveSubscription:*** The invoking entity will be removed and notification about the event would not be dispatched to it.
***Notify:*** Event notifications for which the entity have subscribed will be delivered through this operation.
***Export:*** This operation can trigger a sender initiated form of migration. This operation has the move semantics. The Activity is terminated and started on the remote machine which is specified in the parameters to the operation.
***ExportCopy:*** This operation is similar to the Export operation but it has the copy semantics instead of move. The existing Activity keeps running and a new Activity is started on the specified remote machine.

***Import:*** This is the receiver-initiated form of migration. It has the same semantics as Export but it is triggered by the receiving machine.
***ImportCopy:*** The counterpart of ExportCopy but it takes the receiver-initiated form of migration.

## IV. DGET CUMMUNICATION MODEL

The DGET system has been designed to operate without any centralised server and as such can be regarded as a peer-to-peer system, in contrast to many centralised grid systems, [FOS 97]. This decision was taken to allow the DGET system to be more robust and fault tolerant as well as increasing its capacity to scale. The advantages of the peer to peer topology have been discussed in sections one and two, and are also demonstrated in [TAL 03]. There has been a lot of work in the fields of peer to peer and grid computing, the former focused on file sharing and the latter focusing on centralised systems. We see a combining of the two to produce Grid middleware which is more suited to dynamic environments than current Grid middleware.

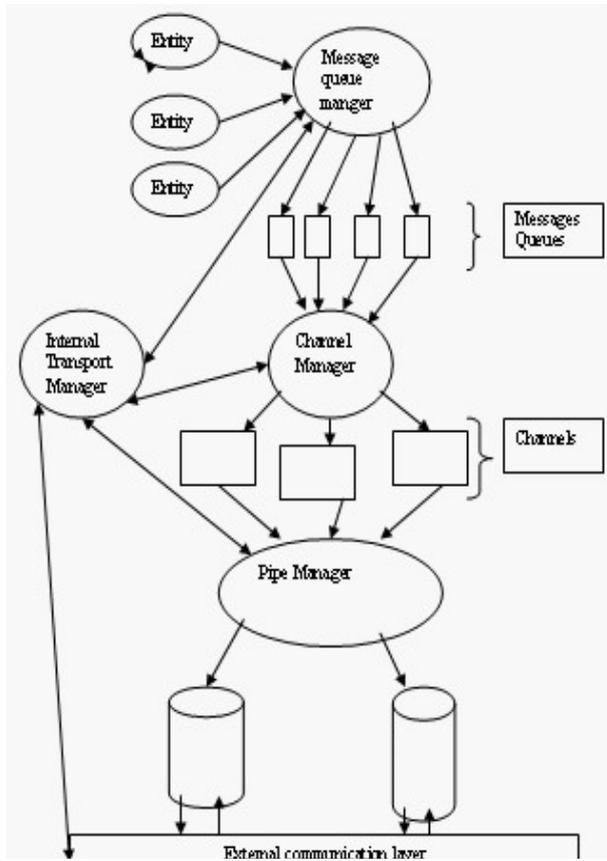

Figure 2: Internal Transport Layer

The communication system is composed of two components, the internal transport layer and the external transport layer of the nucleus.

*1) Internal transport layer*

The internal transport layer is composed of 3 sub layers: message queue, channel and pipe. These layers provide internal abstraction to facilitate various communications operations as shown in [Figure 2].

Entities communicate to each other through connectors, these connectors have message queues, and each one is dedicated to one communication type to one other entity: i.e. incoming, outgoing.

The primary job of a channel is to transform the entity message object to an actual stream of data and also breaking it into small peaces and providing compression, if needed. A channel represents an entity communicating with one or multiple entities through one pipe. A channel is just a multiplexer / de-multiplexer system for the message queue layer. It provides the different operations for breaking down the message into small parts so that it is easier for the pipe to transport them. The inverse operation is provided in the other direction. A channel is always bi-directional.

Another major component of this layer is the delayed channel manager, which provides various methods to handle the migration of entities (manage incoming message during migration and retransmit them etc …). It works a lot like the one from Proactive [BAU 02]. The pipe principally provides a way of communication between two nuclei. It multiplexes / de-multiplexes the messages to/from the various channels it is connected to. It also provides a way of maintaining and negotiating various types of information (security / routing table / …). A pipe is connected to only one external transport protocol through the uniform external communication protocol.

The internal transport layer is the layer that entities see; they use this layer to communicate with each other. If communication needs to be remote they will go through the external communication layer.

*2) External communication layer*

The external communication layer provides an independent interface to various communication protocols, like TCP / UDP / TLS with TCP, HTTP, etc… This layer is here to provide various means of communication between nuclei for transporting messages. Because the network is not totally stable and various obstacles can appear or disappear, we need to use suitable protocols for the various cases that arise. The protocol used between two nuclei or per pipe is negotiated when a pipe is created. That means each nucleus maintains information in their routing table about which protocols other nuclei support.

V. RELATED WORK

DGET can be compared with other solutions on two broader aspects. These are the operational differences of DGET with other solutions and the functional differences. We explain both in the following sections.

A. *Operational Differences*

This category of differences denotes the operational scenarios in which DGET can be deployed. Existing approaches were targeted for a single or a set of operational environments where they can be deployed.

Grid systems were in general built to share high performance computing, storage and specialized scientific instruments among scientists belonging to a few different organizations. The participants of these systems were organizations willing to share their resources with each other. P2P systems mostly focused on sharing file resources where the participants were individual home users [MAT 02, LIA 04]. Other P2P approaches shared the unused CPU power and again the participants were individual PC owners who volunteered their PCs to be used while inactive [AND 02]. DGET contrary to these approaches can be deployed in a wide variety of diverse scenarios. On one extreme it can be used by organizations who want to share their locally controlled organizational resources with other organizations. This deployment scenario is the same in which current grid systems are deployed but DGET has less management and setup overhead compared to other grid solutions. On the other extreme DGET can be easily deployed in a scenario where the participants are individual users who want to share their files or idle CPU power. This aspect is coupled to a functional difference which we explain in the next section.

*B. Functional Differences*

This section will explain what functional differences DGET has compared with other solutions. We explain what components of DGET are different and how they carry out their functionality. As mentioned in the previous section, DGET can be compared to Grid, P2P and hybrid systems. We are going to highlight the difference in three sections devoted to Grid, P2P and hybrid systems.

*1) DGET and Grid Systems:*

In this section we compare DGET with the existing grid solutions. A wide range of systems have been developed. Some of these focus on providing the core middleware services while other programming frameworks are built on top of these middleware systems and provide high-level application development functionalities. Since DGET is a complete solution and besides providing the core middleware services, it has to offer the high level functionality as well, it would be right to compare DGET to both core middleware and high-level programming frameworks individually.

The Globus Toolkit [FOS 97] is the most widely used grid middleware. It is an open source research project developed by Argonne National Laboratories. It has several deployments worldwide. It provides low-level services to the applications and other high-level grid systems. DGET has certain commonalities and a number of differences with the Globus Toolkit. Following are the functional differences with Globus.

The organization of system topology in Globus is manual and static whereas in DGET it is automated. Nodes in the system self-organize themselves dynamically and the topology of the system evolves as a result of interactions in the system. The self organizing mechanism can be adapted to the architecture and needs of DGET users and is not hard coded into the DGET core.

Globus uses hierarchical index services [CZA01] to maintain information about the resources shared in the system. These services need to be manually configured by the system administrators. These index services are queried during the resource discovery process. DGET on the other hand doesn't need to setup any specialized servers. Resource Discovery in DGET is based on the P2P discovery model and thus is totally decentralized. Nodes are organized into an overlay network which takes into consideration user preferences thus increasing the user's perceived efficiency and productivity of the system. Another advantage of adopting a P2P based resource discovery model is scalability issues. Having specialized index servers can be efficient when the number of participants is low but as the number of participants grows to enormous size, performance will start degrading.

On the security front, Globus possess an extremely powerful security system but it has considerable management overhead. All the users are required to have individual accounts on the resource before they can use the resource. This situation is applicable if there are a limited number of participants. In a situation where a very large number of users are present this technique would become very cumbersome. DGET on the contrary doesn't require users to have individual user accounts on the resources. DGET's security mechanism is based on an extended Java security model. Other aspects where DGET security differs from Globus are the fine-grained access control policies and the resource quota control. DGET uses XACML [XACML] to define fine-grained access control policies.

Legion [GRI 97] has the same set of differences with DGET as Globus has. Legion also has a static topology and manual organization of the system components is required. Legion doesn't have a hierarchical system information repository. Information about resources is kept in multiple Collection Objects [CHA 99]. These Collection Objects can suffer from the same scalability drawbacks as the Globus index service. Security in Legion is also very powerful and all the object methods invocations are preceded by a "MayI" method invocation which decides if the invocation should be allowed or not.

UNICORE [ALM 99] (Uniform Interface to COmputing REsources) grid middleware is a German project which allows sharing of computing resources on a wide area scale. UNICORE is not as powerful a system as either

Globus or Legion. System information is static and must be entered by the administrators. Support for dynamic resource discovery is not present. Resources must be specified while submitting jobs. Migration is also not supported.

DGET will also provide job migration while Globus Legion and Unicore do not support it. DGET job entities can be migrated if its resource requirements change, its resource condition changes or as a result of load balancing in the system.

*2) DGET and P2P systems*

The second class of system we can compare DGET with are P2P systems. DGET bears some similarities with P2P systems but uses a significantly different approach in other aspects. We are not going to compare DGET with individual P2P systems rather we will describe the aspects where DGET employs a different approach compared to P2P systems in general.

DGET maintains a decentralized overlay of nodes similar to P2P systems [DRU 01, KUB 00]. DGET does not impose overlay topologies. Various entities can choose to implement the topology that will suit their needs or use the default one provided. DGET adopts a different strategy while building the topology of the system. DGET selects neighbourhood [ELL 04] intelligently taking into consideration the preferences and needs of the nodes. Such customized neighbourhood increases the likelihood of finding resources in the first hop. The neighbourhoods of the nodes evolve as a result of changes in the system.

Most of the P2P systems share file resources only and thus lack sophisticated resource management services. Coordinated use of resources at multiple sites is not supported at present in grid systems. DGET provides powerful resource management entities for a wide variety of resources.

Security is not a major concern in P2P systems therefore most of the P2P systems lack a sophisticated security component. Most of the P2P systems focus on maintaining anonymity [CLA 02]. DGET employs a very sophisticated security mechanism based on the Java security model.

P2P systems lack the support of migration capabilities but DGET is designed to efficiently migrate the entities triggered by a change in the needs of the entities, variation in the system conditions or as a result of load balancing or task scheduling.

*3) DGET and Hybrid Systems*

Some system designers have tried integrating both P2P and Grid approaches to come up with a system which enjoys the benefits of both grid and P2P systems [TAL 03]. This section compares DGET's approach with such hybrid P2P Grid systems. The system we have seen so far is a distributed event brokering system called NaradaBrokering [PAL 03].

NaradaBrokering is a distributed event brokering system which is composed of brokers organized into clusters. These brokers disseminate events to the nodes that have shown interest in such events. NaradaBrokering shares a number of similarities with DGET including a transport independent communication protocol, efficient searching capabilities etc. But on the other hand, NaradaBrokering focuses only on the efficient dissemination of events and doesn't provide any of the solutions provided by the other grid systems including sophisticated and coordinated use of resources located at multiple sites.

## VI. CONCLUSION

In this paper we have presented a peer-to-peer based grid middleware system. The architecture and motivation for the design have been presented. The choices and differences with other systems are also discussed. The GDET project is still in its early stages. So far we have developed a prototype of DGET, which demonstrates many of the system features, such as the entity concept and the peer to peer communication mechanism. The next goal is to finish the implementation and testing of the main components. This will involve adding more fault tolerance and error handling, as well as providing a client interface for the user to administer the system. The testing will involve stressing the system on real-world scientific applications. When the system is more developed it will be released on the web pages of the group and will open to public community for more testing, stressing, and suggestions.